\documentclass[oneside,american]{elsart}
\usepackage[T1]{fontenc}
\usepackage[latin1]{inputenc}
\usepackage{graphicx}
\usepackage{amssymb}

\makeatletter

\providecommand{\boldsymbol}[1]{\mbox{\boldmath $#1$}}

\usepackage{slashed}
\usepackage{color}

\usepackage{babel}
\makeatother
\begin{document}
\begin{frontmatter}

\title{Photon polarization as a probe for quark-gluon plasma dynamics}

\author{Andreas Ipp, Antonino Di Piazza, J\"org Evers, Christoph H.~Keitel}

\ead{andreas.ipp@mpi-hd.mpg.de}

\ead{joerg.evers@mpi-hd.mpg.de}

\address{Max Planck Institute for Nuclear Physics, Saupfercheckweg 1, D-69117
Heidelberg, Germany}

\begin{abstract}
Prospects of measuring polarized photons emitted from a quark-gluon
plasma (QGP) are discussed. In particular, the detection of a possible
quark spin polarization in a QGP using circularly polarized photons
emitted from the plasma is studied. Photons leave the QGP without
further interaction and thus provide a primary probe for quark polarization
within the QGP. We find that photon polarization cannot solely arise
due to a possible QGP momentum space anisotropy, but may be enhanced
due to it. In particular, for oblate momentum distributions and high
photon energies, quark polarization is efficiently transfered to photon
polarization. The role of competing sources of polarized photons in
heavy-ion collisions is discussed. 
\end{abstract}
\end{frontmatter}
The discovery of the quark-gluon plasma (QGP), a new state of matter
at temperatures above $T\gtrsim170$ MeV, in heavy-ion collisions
at CERN/SPS \cite{Heinz:2000bk} and RHIC \cite{Gyulassy:2004zy}
leaves open many questions about its properties and its internal structure:
If the naive picture of free quarks and gluons is obsolete, is the
QGP rather a strongly coupled system \cite{Shuryak:2004cy}, or more
adequately described as a fluid with low viscosity \cite{Teaney:2003kp}?
How does the QGP apparently thermalize so fast ($\lesssim1$ fm/c)
\cite{Arnold:2004ti}? Are quarks in the QGP polarized \cite{Liang:2004ph}?
In order to learn more about the QGP, one would like to have a primary
probe directly from its interior, which is difficult due to its very
short lifetime $\tau\sim5$ fm/c. Such a probe is given by direct
thermal photons that, when produced, typically escape the plasma without
further interaction. The PHENIX experiment at RHIC has confirmed the
good agreement between the photon rate measured and perturbative calculations
\cite{Adler:2005ig}. It has been suggested that direct photons could
provide information about a possible momentum anisotropy within the
QGP \cite{Schenke:2006yp}. Polarized photon emission from the QGP
has been previously studied as a geometrical surface effect \cite{Goloviznin:1988ke},
but to our knowledge not from the bulk.

Global quark polarization within the QGP has been proposed by Liang
and Wang \cite{Liang:2004ph} to occur in non-central heavy-ion collisions
by a simple mechanism: Due to a finite impact parameter, the bulk
of the QGP is kicked to rotate within the reaction plane, and the
resulting potentially large orbital angular momentum may lead to quark
spin polarization due to spin-orbit coupling (see also \cite{Voloshin:2004ha,Liang:2004xn}).
A possible transfer of quark polarization on massive particles was
studied for the case of $\Lambda$-hyperons. Experimentally, an upper
limit of the polarization $|P_{\Lambda,\bar{\Lambda}}|\lesssim0.02$
has been obtained at RHIC \cite{Selyuzhenkov:2006combined,Chen:2007zzq},
while recent calculations suggest $P_{\Lambda}\approx-0.05$ \cite{Liang:2007ma,Gao:2007bc}.
It should be noted though that hyperon polarization is affected by
all stages of the collision, including the later hadronization and
the interacting hadron gas phase, to an unknown degree \cite{Betz:2007kg}. 

In this Letter, we show that global QGP polarization would effectively
lead to a polarization of photons. Photons are a primary probe as
they are likely to leave the plasma without further interaction. Both
angular distribution and polarization of the emitted photons depend
on the quark polarization. Particularly, we point out the possibility
to detect a possible global quark polarization using circularly polarized
photons. We also show that a momentum-anisotropy alone does not lead
to a polarization of the emitted photons. Therefore, the polarization
of photons can be facilitated as a direct signal of the quark spin
polarization.

Thermal photons in the QGP can be produced through Compton scattering
of (anti-)quarks and gluons, $qg\rightarrow q\gamma$ ($\bar{q}g\rightarrow\bar{q}\gamma$),
and annihilation of quarks and antiquarks, $q\bar{q}\rightarrow g\gamma$.
The bulk of the QGP is dominated by up and down quarks, for which
the ultrarelativistic limit $T\gg m$ for the current quark masses
$m$ applies. In thermal equilibrium, photons are emitted isotropically,
thermally distributed, and unpolarized. In a heavy-ion collision,
however, the QGP expands along the beam axis which induces momentum
space anisotropy, and the produced photons may reflect this in their
rapidity dependence \cite{Schenke:2006yp}. For clarity, we will use
the term {}``anisotropy'' for momentum space anisotropy within the
QGP, {}``polarization'' for quark spin or photon polarization, and
we use the temperature $T$ as hard energy scale ($q_{{\rm hard}}=T$),
even though it is properly defined only in the isotropic case. Also,
$\hbar=c=k_{B}=1$.

The spin of a quark with momentum $\kappa=(\kappa_{0},\boldsymbol{\kappa})$
can be defined in its rest frame \cite{Bjorken:1965aa}. Under a boost
along $\boldsymbol{\kappa}$, only the longitudinal component of the
spin four-vector $s_{{\rm rest}}=(0,\mathbf{s}_{{\rm rest}})$ changes,
the transverse contribution being invariant. For an ensemble of quarks
with momentum $\boldsymbol{\kappa}$, the average spin $\mathbf{s}_{{\rm rest}}$
can be split according to $\mathbf{s}_{{\rm rest}}=p^{\parallel}\hat{\boldsymbol{\kappa}}+p^{\perp}\hat{\mathbf{s}}^{\perp}$
\cite{Olsen:1980cw}, where $p^{\parallel}=p^{\parallel}(\hat{\mathbf{\boldsymbol{\kappa}}})$
and $p^{\perp}=p^{\perp}(\hat{\boldsymbol{\kappa}})$ denote the degree
of polarization along the propagation direction $\hat{\boldsymbol{\kappa}}\equiv\boldsymbol{\kappa}/|\boldsymbol{\kappa}|$
or along a transverse direction $\hat{\mathbf{s}}^{\perp}=\hat{\mathbf{s}}^{\perp}(\hat{\boldsymbol{\kappa}})$
with $\hat{\boldsymbol{\kappa}}\cdot\hat{\mathbf{s}}^{\perp}=0$.
States with partial polarization can be expressed in the ultrarelativistic
limit using the density matrices $\rho_{\pm}=\slashed{\kappa}P_{\pm}$
with $P_{\pm}=\left(1\mp p^{\parallel}\gamma^{5}+p^{\perp}\gamma^{5}\slashed{\hat{s}}^{\perp}\right)/2$,
where the upper (lower) sign refers to particles (antiparticles) \cite{Bjorken:1965aa}.
Partial polarization $|\mathbf{s}_{{\rm rest}}|^{2}<1$ is understood
as a linear combination $p^{\parallel}=p^{\parallel}(\uparrow)-p^{\parallel}(\downarrow)$
with $p^{\parallel}(\uparrow)+p^{\parallel}(\downarrow)=1$ (and likewise
for $p^{\perp}$) of fully polarized contributions $|\mathbf{s}_{{\rm rest}}|^{2}=1$,
for which $P_{\pm}^{2}=P_{\pm}$ are projectors \cite{Bjorken:1965aa}. 

\begin{figure}
\begin{center}\includegraphics[%
  scale=0.7]{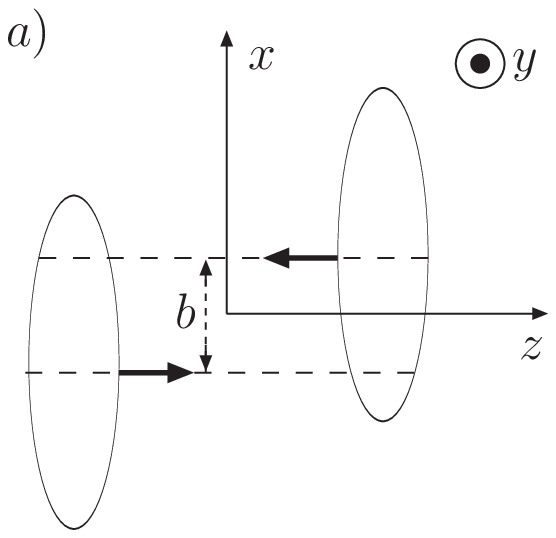}~~~~~~~\includegraphics[%
  bb=206bp 561bp 389bp 721bp,
  scale=0.7]{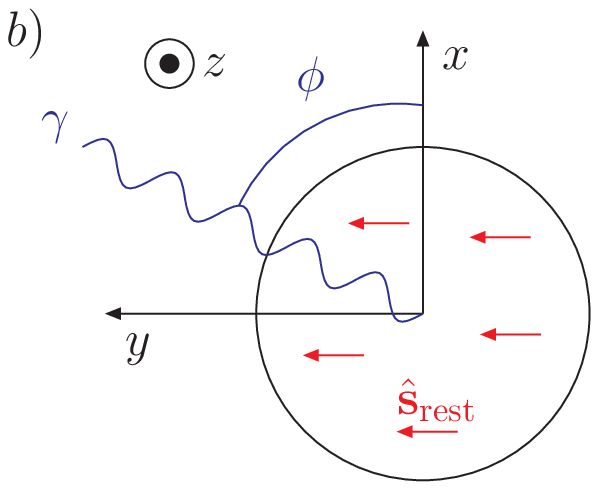}\end{center}

\caption{(Color online). (a) Heavy ion collisions with non-zero impact parameter
$b$ may produce a global spin polarization of quarks along $\hat{\mathbf{y}}$
as proposed in Ref.~\cite{Liang:2004ph}. (b) Such a global quark
spin $\hat{\mathbf{s}}_{{\rm rest}}=\hat{\mathbf{y}}$ could be detected
through the azimuthal $\phi$-dependence of photons emitted from the
plasma. \label{fig:globalspin}}
\end{figure}

In principle, for a certain point in coordinate space, momentum dependent
polarization functions $p^{\parallel}(\kappa)$, $p^{\perp}(\kappa)$,
and $\hat{s}^{\perp}(\kappa)$ for each (anti-)quark species would
completely specify the quark polarization content within the QGP.
Ideally, these functions should follow from some first-principle calculation
in the medium. Since they are unknown, we proceed in a pragmatic way:
in accordance with global polarization proposed for non-central collisions
\cite{Liang:2004ph,Gao:2007bc} (see Fig.~\ref{fig:globalspin}),
we assume that the spin of each quark in its rest frame is aligned
along the same direction $\hat{\mathbf{s}}_{{\rm rest}}(\kappa)=\hat{\mathbf{y}}$
and that particles with same energy $\kappa_{0}$ share the same degree
of polarization $p_{{\rm rest}}(\kappa_{0})=\left|\mathbf{s}_{{\rm rest}}(\kappa_{0},\boldsymbol{\kappa})\right|$.
Since the main polarization effect is expected to arise from hard
modes with energy $\kappa_{0}\sim T$ \cite{Liang:2007ma,Gao:2007bc},
we use the following model for the energy dependence of $p_{{\rm rest}}(\kappa_{0})$:
\begin{equation}
p_{{\rm rest}}(\kappa_{0})=\bar{p}_{{\rm rest}}\Theta(\kappa_{0}-k^{*}).\label{eq:polarizationmodel}\end{equation}
The energy threshold $k^{*}$ is chosen equal to the separation scale
between soft and hard modes, to be specified below Eq.~(\ref{eq:comptonspinpolarized2}).
Modes that greatly exceed the temperature are automatically cut off
by the thermal distribution functions of the plasma. 

\begin{figure}
\begin{center}\includegraphics[%
  scale=0.7]{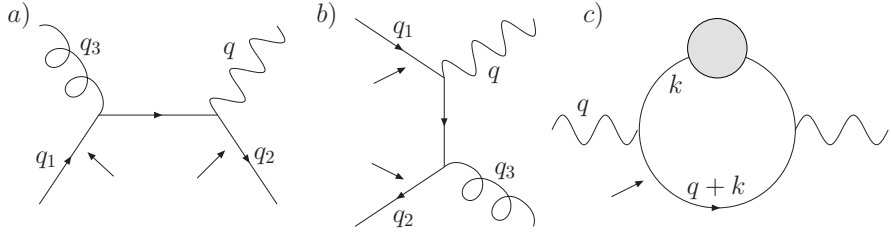}\end{center}

\caption{Feynman diagrams corresponding to (a) Compton scattering, (b) quark-antiquark
annihilation, and (c) photon self-energy (straight, curly, and wiggly
lines represent quarks, gluons, and photons). Arrows indicate where
polarized states are used in the calculation. The gray filled circle
indicates a soft propagator dressed by the medium. \label{fig:diagrams}}
\end{figure}

The calculation of the production rate of photons with momentum $q=(E,\mathbf{q})$
requires a separation into hard ($\sim T$) and soft ($\sim gT$)
momentum transfer contributions \cite{Kapusta:2006,Schenke:2006yp}.
We restrict ourselves to hard $2\leftrightarrow2$ particle processes
and corresponding soft processes, and will not consider processes
involving more particles, like bremsstrahlung and inelastic pair annihilation.
A consistent treatment of the latter processes requires an analysis
of multiple soft scattering \cite{Arnold:2001ms}, which in the anisotropic
case still poses fundamental open questions due to the possibility
of plasma instabilities \cite{Schenke:2006yp,Arnold:2004ti}. In the
isotropic case, these processes change the total photon rate in the
relevant energy range $2.5\leq E/T\leq10$ by a factor of two \cite{Arnold:2001ms}.
For hard momenta, standard perturbation theory is used to calculate
the rate contributions from Compton scattering $R_{{\rm c}}$ or annihilation
$R_{{\rm a}}$ \begin{eqnarray}
E\frac{dR_{{\rm c}}}{d^{3}q} & = & 20\pi\int_{\delta}f_{\xi}^{F}(\mathbf{q}_{1})f_{\xi}^{B}(\mathbf{q}_{3})\left(|{\cal M}_{{\rm c}}^{S}|^{2}\!-\! f_{\xi}^{F}(\mathbf{q}_{2})|{\cal M}_{{\rm c}}|^{2}\right),\nonumber \\
E\frac{dR_{{\rm a}}}{d^{3}q} & = & \frac{320\pi}{3}\int_{\delta}f_{\xi}^{F}(\mathbf{q}_{1})f_{\xi}^{F}(\mathbf{q}_{2})\left(1+f_{\xi}^{B}(\mathbf{q}_{3})\right)|{\cal M}_{{\rm a}}|^{2},\label{eq:photonproductionratehard}\end{eqnarray}
with $\int_{\delta}\equiv\int d^{3}q_{1}d^{3}q_{2}d^{3}q_{3}/(2\pi)^{9}\delta^{(4)}(q_{1}+q_{2}-q_{3}-q)/(8E_{1}E_{2}E_{3})$
(the momenta are defined in Fig.~\ref{fig:diagrams}), and fermionic
or bosonic distribution functions $f_{\xi}^{F/B}$. The numerical
prefactors include summing over $N_{c}=3$ colors and the fractional
charge factors of the considered up and down quarks that dominate
the rate due to their low mass. Pauli blocking has to be taken into
account properly by including the scattering Compton matrix element
either polarized (${\cal M}_{{\rm c}}$), or summed over outgoing
quark polarization (${\cal M}_{{\rm c}}^{S}$). For anisotropic momentum
distributions, the isotropic distribution functions $f_{{\rm iso}}^{F/B}$
are assumed to be stretched along the beam axis $\mathbf{z}$ in momentum
space \cite{Romatschke:2003ms}, $f_{\xi}^{F/B}(\mathbf{k})=N(\xi)f_{{\rm iso}}^{F/B}\left(k_{\xi}\right)$,
with $k_{\xi}^{2}=k_{x}^{2}+k_{y}^{2}+(1+\xi)k_{z}^{2}$, the anisotropy
parameter $\xi>-1$, and a normalization factor which we choose as
$N(\xi)=1$ as in \cite{Schenke:2006yp}. A different normalization
is discussed later. 

Before looking at globally polarized quarks, we first want to study
the polarization of emitted photons from a QGP without quark polarization.
In the following, we present the result for the Compton scattering
diagram Fig.~\ref{fig:diagrams}(a), since the corresponding scattering
matrix ${\cal M}_{{\rm a}}$ for the annihilation diagram Fig.~\ref{fig:diagrams}(b)
follows from crossing symmetry. For polarized Compton scattering with
electron charge $e$ and strong coupling $g$ we find\begin{equation}
|{\cal M}_{{\rm c}}|^{2}=\frac{e^{2}g^{2}}{6}\left(\frac{q\cdot q_{1}}{q\cdot q_{2}}+\frac{q\cdot q_{2}}{q\cdot q_{1}}-2m^{2}\left|\frac{\epsilon\cdot q_{1}}{q\cdot q_{1}}\!-\!\frac{\epsilon\cdot q_{2}}{q\cdot q_{2}}\right|^{2}\right).\label{eq:comptonpolarized}\end{equation}
In this equation we have kept a finite quark mass $m$ dependence
to demonstrate how the polarization vector $\epsilon(q)$ would enter
the cross section. This formula leads to the well-known Klein-Nishina
differential cross section \cite{Tashenov:2006a}. For isotropic momentum
distributions any polarization effect averages out. However, for anisotropic
momentum distributions, finite quark masses may give rise to linear
polarization of photons, even if the quarks in the plasma are unpolarized.
But in the ultrarelativistic limit considered, Eq.~(\ref{eq:comptonpolarized})
trivially reduces to the unpolarized case. \underbar{}For \underbar{}the
soft part, the dressed quarks in the medium acquire a thermal mass
of the parametric order $gT$. If this thermal mass adopts the role
of an effective quark mass, photon polarization in an anisotropic
medium would not yet be excluded. To address this question, we calculate
the soft contributions for unpolarized quarks and polarized photons.
These can be \underbar{}obtained from the imaginary part of the photon
self-energy, which corresponds to the 12-component of the photon self-energy
$\Pi$ in Fig.~\ref{fig:diagrams}(c) in the real-time formalism
\cite{Kapusta:2006}\begin{equation}
E\frac{dR_{{\rm soft}}}{d^{3}q}=-\frac{i}{2(2\pi)^{3}}\epsilon_{\mu}^{*}\epsilon_{\nu}\Pi_{12}^{\mu\nu}(q).\label{eq:photonratesoftpolarized}\end{equation}
The unpolarized photon form \cite{Baier:1997xc,Schenke:2006yp} is
recovered by applying the Ward identity. For general polarization
vectors $\epsilon(q)$ one finds within the Hard Loop approximation
(which assumes $gT\sim k\ll q\sim T$) \cite{Kapusta:2006}\begin{eqnarray}
i\epsilon_{\mu}^{*}\epsilon_{\nu}\Pi_{12}^{\mu\nu}(q) & = & e^{2}\frac{20f_{\xi}^{F}(\mathbf{q})}{3|\mathbf{q}|}\int\frac{d^{3}k}{(2\pi)^{3}}{\rm Im}\, q\cdot S_{R}^{\star}(k)\label{eq:Pisoftpolarized}\end{eqnarray}
with the retarded dressed fermionic propagator $\slashed{S}{}_{R}^{\star}(k)$
\cite{Schenke:2006fz}. This result is independent of the polarization
vector and is just 1/2 of the unpolarized result \cite{Baier:1997xc,Schenke:2006yp}.
This means that within our model, photons emitted from a QGP without
quark polarization are unpolarized, even if the distribution functions
show momentum anisotropy. In other words, the quarks themselves have
to be polarized in order to observe photon polarization, and we can
directly relate photon polarization to quark polarization.

We analyze the quark spin dependence by using polarized states $\rho_{\pm}$
as indicated in Fig.~\ref{fig:diagrams}, which extends Eq.~(\ref{eq:comptonpolarized})
in the ultrarelativistic case to \begin{eqnarray}
|{\cal M}_{{\rm c}}|^{2} & = & \frac{e^{2}g^{2}}{12}\left\{ \left(1+p^{\parallel}(q_{1})p^{\parallel}(q_{2})\right)\left(\frac{q\cdot q_{1}}{q\cdot q_{2}}+\frac{q\cdot q_{2}}{q\cdot q_{1}}\right)\right.\label{eq:comptonspinpolarized2}\\
 &  & +\left(p^{\parallel}(q_{1})+p^{\parallel}(q_{2})\right)\left(\frac{q\cdot q_{2}}{q\cdot q_{1}}-\frac{q\cdot q_{1}}{q\cdot q_{2}}\right)i{\rm det}\left|\boldsymbol{\epsilon}\boldsymbol{\epsilon}^{*}\hat{\mathbf{q}}\right|\nonumber \\
 &  & -2p^{\perp}(q_{1})p^{\perp}(q_{2})\left[\left(q\cdot\hat{s}_{1}^{\perp}\right)\left(q\cdot\hat{s}_{2}^{\perp}\right)\left(\frac{1}{q\cdot q_{1}}-\frac{1}{q\cdot q_{2}}\right)\right.\nonumber \\
 &  & \left.\left.+\hat{s}_{1}^{\perp}\cdot\hat{s}_{2}^{\perp}-\frac{\left(q\cdot\hat{s}_{1}^{\perp}\right)\left(q_{1}\cdot\hat{s}_{2}^{\perp}\right)}{q\cdot q_{1}}-\frac{\left(q\cdot\hat{s}_{2}^{\perp}\right)\left(q_{2}\cdot\hat{s}_{1}^{\perp}\right)}{q\cdot q_{2}}\right]\right\} \nonumber \end{eqnarray}
with $\hat{s}_{1}^{\perp}=\hat{s}^{\perp}(q_{1})$ and $\hat{s}_{2}^{\perp}=\hat{s}^{\perp}(q_{2})$.
For unpolarized outgoing quarks one has $|{\cal M}_{{\rm c}}^{S}|^{2}=2|{\cal M}_{{\rm c}}|^{2}$,
with $p^{\parallel}(q_{2})=p^{\perp}(q_{2})=0$. The pieces quadratic
in the polarization, proportional to $p^{\parallel}(q_{1})p^{\parallel}(q_{2})$
and $p^{\perp}(q_{1})p^{\perp}(q_{2})$, vanish independently in the
infrared (IR) limit through the angular integrations in (\ref{eq:photonproductionratehard}),
but they may still contribute at larger momentum transfers as we find
numerically. Only the piece linear in the polarization, proportional
to $p^{\parallel}(q_{1})+p^{\parallel}(q_{2})$, survives in the IR
limit. This piece has a matching ultraviolet contribution from the
soft sector: Due to the assumption in Eq.~(\ref{eq:polarizationmodel})
we only need to take into account polarized states for the hard quark
line in Fig \ref{fig:diagrams}(c). 

For polarized quarks and photons, Eq.~(\ref{eq:Pisoftpolarized})
is extended to\begin{eqnarray}
i\epsilon_{\mu}^{*}\epsilon_{\nu}\Pi_{12}^{\mu\nu}(q) & = & \left(1+p^{\parallel}(q)i{\rm det}\left|\boldsymbol{\epsilon}\boldsymbol{\epsilon}^{*}\hat{\mathbf{q}}\right|\right)\quad\label{eq:Pisoftpolarized2}\\
 &  & \times e^{2}\frac{10f_{\xi}^{F}(\mathbf{q})}{3|\mathbf{q}|}\int\frac{d^{3}k}{(2\pi)^{3}}{\rm Im}\, q\cdot S_{R}^{\star}(k).\nonumber \end{eqnarray}
The determinant in this expression also occurs in the hard part in
Eq.~(\ref{eq:comptonspinpolarized2}). It vanishes for linear polarization,
but contributes for right-(left)-handed circular polarization with
$i\det\left|\boldsymbol{\epsilon}\boldsymbol{\epsilon}^{*}\hat{\mathbf{q}}\right|=+(-)1$.
The photon rate $EdR/d^{3}q$ is finally given by adding the results
of Eq.~(\ref{eq:photonproductionratehard}) and (\ref{eq:photonratesoftpolarized}),
with an intermediate cutoff $k^{*}\sim\sqrt{g}T$ that is varied to
estimate uncertainties due to its choice, using Monte Carlo integration
\cite{Schenke:2006yp}.

\begin{figure}
\begin{center}\includegraphics[%
  scale=0.9]{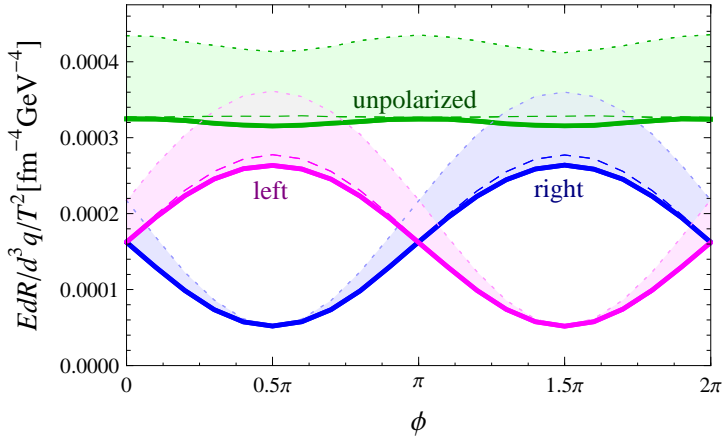}\end{center}

\caption{(Color online). Photon rate for $E/T=5$ and $\xi=0$ for full quark
polarization $\bar{p}_{{\rm rest}}=1$, separated in left and right
circularly polarized and unpolarized photons, as a function of the
azimuthal angle $\phi$ in the $x$-$y$-plane. The hard-soft separation
scale is varied by a factor 2 up (dashed line) or down (dotted line)
around its central value (thick line). \label{fig:polarized}}
\end{figure}

As a typical example, Fig.~\ref{fig:polarized} shows the photon
rate for isotropic momentum distribution with $\xi=0$ as a function
of the angle $\phi$ in the $x$-$y$-plane (see Fig.~\ref{fig:globalspin}),
separated in the two circularly polarized states and their unpolarized
sum. Results for other $\xi$ values look qualitatively similar. For
the QGP where the hard energy scale exceeds $T\gtrsim200\,{\rm MeV}$,
the ratio $E/T=5$ corresponds to photon energies of $1\,{\rm GeV}$
or higher. Photons with such high energy are emitted during the initial
stage and are not affected by the subsequent expansion. Background
effects typically set in at lower energies. This is complementary
to the studies in \cite{Goloviznin:1988ke} where much lower photon
energies were considered. One observes a maximum for left (right)
circular polarized photons along (opposite to) the direction of the
global spin. This is in accordance with high-energy Compton scattering,
where helicity states couple to circularly polarized photons. There
appears to be some angular dependence also for the unpolarized photon
rate \cite{Adler:2005rg,Turbide:2005bz,Chatterjee:2005de}. We verified
that this effect is physical for prolate momentum anisotropies with
$\xi\approx-1$ (corresponding to early stages of the collision) where
it arises from the angular dependence of unpolarized photon emission
from quark-antiquark annihilation with transverse spin. For isotropic
and oblate systems in momentum space with $\xi\gtrsim0$, the unpolarized
photon rate only shows small $\phi$-dependence such that the non-negligible
dependence of the unpolarized effect on the choice of the intermediate
cutoff scale $k^{*}$ prevents definitive conclusions at this point.
Apart from circular polarization, the remaining Stokes parameters
that describe linear polarization all vanish.

\begin{figure}
\begin{center}\includegraphics[%
  scale=0.9]{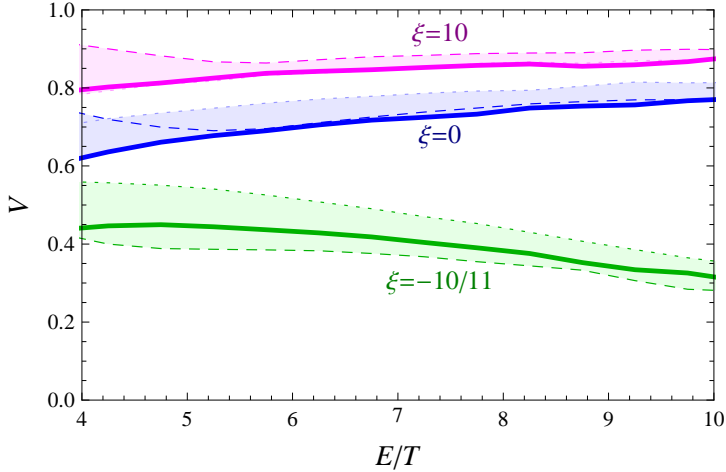}\end{center}

\caption{(Color online). Visibility $V$ of photons with circular polarization
as a function of photon energy for various anisotropy parameters $\xi$.
Bands mean the same as in Fig.~\ref{fig:polarized}. \label{fig:visibility}}
\end{figure}

From the maximum (max) and minimum (min) photon rate as a function
of $\phi$ we obtain the visibility $V=\left({\rm max}-{\rm min}\right)/\left({\rm max}+{\rm min}\right)$.
Figure \ref{fig:visibility} depicts $V$ for polarized photons as
a function of the photon energy. In this ratio, systematic uncertainties
are largely cancelled. We find that the visibility increases with
growing oblate anisotropy $\xi\gtrsim0$ (corresponding to later stages
of the collision) and photon energy, but decreases for prolate distributions
$\xi<0$. For a different choice of the prefactor $N(\xi)=\sqrt{1+\xi}$
for the anisotropy distribution functions used in \cite{Schenke:2006fz}
one finds a similar dependence on $\xi$, but less pronounced. For
energies $E/T\lesssim4$, the influence of our assumption (\ref{eq:polarizationmodel})
cannot be neglected, but for larger energies, the result appears to
be rather insensitive to the choice of the intermediate cutoff scale
$k^{*}$. The visibility is proportional to global polarization $\propto\bar{p}_{{\rm rest}}$
(e.g.~for $\xi=10$ and global spin polarization $\bar{p}_{{\rm rest}}\approx0.05$,
the visibility would be $V\approx0.04$), and is largest in the $x$-$y$-plane.
Thus, the quark polarization is imprinted on emitted photons. Therefore
photon polarization can be used to measure spin polarization, provided
one can detect the polarization state of high-energy photons.

Due to the involved dynamics in heavy-ion collisions, there is a rich
variety of possible photons sources. One usually separates the contributions
into direct photons, fragmentation photons, and background photons
\cite{Reygers:2002kc,Turbide:2005fk}. Direct photons are those produced
through parton collisions, like Compton scattering and quark - antiquark
annihilation. Apart from thermal photons from the quark-gluon plasma,
direct photons also include prompt photons from initial hard parton
scatterings, as well as thermal photons from a later hadron gas phase.
Fragmentation photons are produced by bremsstrahlung emitted from
final state partons. Background photons are produced by hadron decay
in later stages of the collision such as $\pi^{0}\rightarrow\gamma\gamma$
or $\eta\rightarrow\gamma\gamma$. Additionally, high-energy partons
in the form of jets can produce photons through direct interaction
with the medium and through in-medium bremsstrahlung. In the following,
we will analyze to what extent these processes could lead to the production
of polarized photons. 

Prompt photons dominate the photon spectrum at the highest energies.
If one considers the leading-order contributions from Compton scattering
and annihilation, one sees from Eq.~(\ref{eq:comptonspinpolarized2})
that the photons produced can only be polarized if the incoming quarks
are polarized. This could occur through spin polarization of the ions
in the storage ring. The polarization of incoming quarks could also
be triggered through a strong magnetic field that is created in non-central
collisions as explained below. However, this energy range is not of
interest for thermal photon production, because at this energy range
thermal photons are highly suppressed.

For energies below a few GeV, background photons become increasingly
dominating and difficult to subtract \cite{Reygers:2002kc}. The main
contribution of background photons originates from $\pi^{0}\rightarrow2\gamma$
decay. But since the pion has spin zero \cite{Brommel:2005ee,Yang:1950rg},
it can not carry any spin information out of the plasma, and thus
no net polarization can be transfered to the two photons produced.
Thus, background photons from $\pi^{0}$ do not introduce additional
photon polarization.

Experimentally, the energy region of a few GeV is the most promising
to look for thermal photons. Here photons produced through jets compete
with or even exceed the prompt photon yield \cite{Turbide:2005fk}.
Most important are photons that are produced through conversion of
a jet in the plasma through Compton scattering or annihilation. Polarized
photons could be produced either if the thermal quark from the QGP
is polarized or if a quark in the jet is polarized. In the former
case a global quark polarization could be reflected in the photons
emitted. In the latter case, strong magnetic fields could lead to
a polarization of quarks and thus to the emission of polarized photons.
It has been pointed out recently that in non-central heavy-ion collisions
indeed the charged remnants of the incoming nuclei that do not participate
in the collision induce a very strong magnetic field in the collision
region that can exceed the critical magnetic field of electrons for
a short time \cite{Kharzeev:2007jp}. This field would lead to a spin
orientation of produced quarks in jets which in turn could be converted
into polarized photons through interaction with the hot medium. A
strong magnetic field would further allow for magnetic bremsstrahlung
\cite{Erber:1966vv}, that is the emission of a photon from a quark
or antiquark in the presence of a background magnetic field - a process
that is kinematically forbidden for a vanishing magnetic field. A
careful quantitative study of all these different possible contributions
is necessary to estimate the experimental detectability of the polarization
produced in the quark-gluon plasma. 

As unpolarized photons are routinely observed in high-energy collisions
\cite{Adler:2005ig,Adler:2005rg}, a main experimental challenge is
to detect circularly polarized photons in the GeV energy range \cite{Avakian:2006xx,Uggerhoj:2005ms}.
We propose to use a single aligned crystal acting as a quarter wave
plate for high energy photons \cite{Apyan:2005dx,Cabibbo:1962a} to
convert their circular polarization into linear polarization that
can then be analyzed by means of another aligned crystal through coherent
electron-positron pair production \cite{Apyan:2005dx,Uggerhoj:2005ms,Cabibbo:1962b}.
The low photon fluxes of the order of a photon per collision that
are expected from the QGP in the GeV energy range would pose an additional
technical difficulty for the experimental detectability. On the other
hand, even if the polarization detector covers only a fraction of
the $4\pi$ coverage at some fixed position in the laboratory, angular
dependence studies can be performed as the reconstructible reaction
plane and thus the global polarization changes its orientation from
event to event.

Concluding, we have shown that \emph{}spin polarization in the QGP
would lead to the emission of circularly polarized photons. \emph{}Good
photon polarization visibility is found in particular for higher photon
energies and oblate momentum distributions. We further showed that
for the leading order processes considered, momentum anisotropy alone
cannot give rise to polarization of photons. With the ongoing experimental
progress in the detection of polarized photons in the GeV range, further
exploration of other sources of polarized photons in heavy-ion collisions
would certainly be desirable.

We thank A.~Apyan, K.~Z.~Hatsagortsyan, Z.-T.~Liang, B.~Schenke,
M.~Strickland, A.~Surzhykov, S.~Tashenov, and U.~I.~Uggerh{\o}j
for helpful discussions.

\bibliographystyle{elsart-num}
\bibliography{polarized_bib}

\end{document}